\documentstyle[amssymb,preprint,aps]{revtex}

\begin{document}
\title{Resonant dynamics in boson-fermion mixtures}
\author{M. Wouters, J. Tempere$^{\ast }$, J. T. Devreese$^{\ast \ast }$}
\address{TFVS, Universiteit Antwerpen, Universiteitsplein 1, B2610 Antwerpen, Belgium.}
\date{January 31, 2003}
\maketitle

\begin{abstract}
The time evolution of a dilute atomic boson-fermion mixture is studied at
zero temperature, when the external magnetic field is brought close to a
interspecies Feshbach resonance. On short time scales we find that before a
collapse or phase separation of the mixture can occur, oscillations between
fermionic atoms and heteronuclear molecules dominate the dynamics. Applying
the model presented here to a trapped mixture of $^{40}$K-$^{87}$Rb atoms,
we find that the predicted oscillations can be revealed in degenerate
mixtures of these atoms.
\end{abstract}


\section{Introduction}

\tighten One of the most attractive features of cold atomic gases is that
the interatomic scattering length is under very good experimental control.
Using the presence of Feshbach resonances, the scattering length can be
tuned from highly repulsive to strongly attractive by changing the applied
magnetic field. Sympathetic cooling techniques allow to cool mixtures of
fermionic and bosonic atoms down to extremely low temperatures where the
quantum mechanical nature of both bosons and fermions becomes apparent,
resulting in a Bose-Einstein condensate coexisting with a filled Fermi sea 
\cite{truscottSCI291,schreckPRL87}. When the interspecies interactions are
strongly attractive, changing the number of bosons alters the mean-field
interaction strength and can result in a collapse of the Fermi gas as
observed by Modugno {\it et al.} \cite{modugnoSCI297}. This instability can
also be triggered by changing the scattering length itself, making use of a
Feshbach resonance. The divergence of the scattering length at a Feshbach
resonance is due to a magnetically tunable molecular state whose binding
energy vanishes at resonance. This means that close to this magnetic field
value, the molecular state can be populated. A number of experimental \cite
{DonleyNature417} and theoretical works \cite
{KokkelmansPRL89,Koehlercond-mat/0209,Duinecondmat0210,MackiePRL89} have
been devoted to study the conversion of atoms between the atomic and
molecular condensates and the thermal cloud.

In this work we study the evolution of an initially noninteracting
boson-fermion mixture at zero temperature when a sudden change (on a time
scale much faster than any other time scale in the system) in the magnetic
field brings it close to an interspecies Feshbach resonance. The (fermionic)
molecules first form out of a condensate boson and a fermionic atom with an
energy smaller than or equal to the Fermi energy. We will refer to such
fermionic atoms (molecules) as `{\it slow}' fermions{\bf \ }(molecules); and
to fermions or molecules with more energy than the Fermi energy as `{\it fast%
}' fermions or molecules, respectively. The decay of a molecule can then
either lead back to a condensate boson with the same `slow' fermion or\ to a
thermal boson accompanied by a `fast' fermion.{\Large \ }Energy conservation
is not violated because the interaction energy is changed. The latter
process shows that the stepwise change of the magnetic field can lead to
heating of the system, while the former demonstrates that coherent
oscillations between fermionic atoms and molecules are possible, when a
fermionic atom interacts with the condensate as a whole. Due to multiple
scattering of fermionic atoms, bosons and molecules, the time evolution of
the `fast' part of the system is expected to be rather chaotic. This is why
we preferred to focus on the `slow' fermionic atoms and molecules. The
energies involved in the formation of molecules are easily two orders of
magnitude bigger than the Fermi energy and thus we conclude from the
difference in phase space volume that it is rather unlikely that the `fast'
decay products of a molecule will come back in the Fermi sphere. This
assumption allows us to study the `slow' fermionic atoms and molecules as a
subsystem.

Our study is based on the model Hamiltonian introduced in Sec. II. We make
some approximations that allow us to write the time dependence of `slow'
fermionic atoms and molecules in analytical form.\ The results obtained with
this simplified model are discussed in Sec. III, together with the
experimental conditions needed to make the predictions observable, using
data from $^{40}$K-$^{87}$Rb scattering.

\section{Description of the dynamics}

Experimentally, the scattering length as a function of the magnetic field
close to a Feshbach resonance can be well parameterized by

\begin{equation}
a_{\text{scatt}}\left( B\right) =a_{bg}\left( 1-\frac{\Delta B}{B-B_{0}}%
\right) ,  \label{ascatt}
\end{equation}
where $a_{bg}$ is the background scattering length, $B_{0}$ the position and 
$\Delta B$ the width of the resonance.

A good description of the scattering physics near a Feshbach resonance can
be obtained by explicitly introducing molecules \cite
{TimmermansPRL83,kokkelmansPRA65}. These molecular states are the bound
states of an interatomic potential in a closed channel and their energy
relative to the threshold is given by 
\begin{equation}
\nu \left( B\right) =\left( B-B_{0}\right) \Delta \mu ,  \label{nu}
\end{equation}
where $\Delta \mu =\mu ^{\text{molecule}}-\mu ^{\text{fermion}}-\mu ^{\text{%
boson}}$ is the difference in magnetic moment between the molecular and
atomic states. Denoting the coupling strength between atoms and molecules by 
$g_{m}$, the Hamiltonian for a trapped gas can be written as 
\begin{eqnarray}
H=\int d^{3}{\bf x\,} &&\left[ \psi _{f}^{\dag }\left( {\bf x}\right)
H_{f}^{0}\left( {\bf x}\right) \psi _{f}\left( {\bf x}\right) +\psi
_{b}^{\dag }\left( {\bf x}\right) H_{b}^{0}\left( {\bf x}\right) \psi
_{b}\left( {\bf x}\right) +\psi _{m}^{\dag }\left( {\bf x}\right)
H_{m}^{0}\left( {\bf x}\right) \psi _{m}\left( {\bf x}\right) \right. 
\nonumber \\
&&\left. +g_{m}\,\psi _{m}^{\dag }\left( {\bf x}\right) \psi _{f}\left( {\bf %
x}\right) \psi _{b}\left( {\bf x}\right) +g_{m}^{\ast }\,\psi _{b}^{\dag
}\left( {\bf x}\right) \psi _{f}^{\dag }\left( {\bf x}\right) \psi
_{m}\left( {\bf x}\right) \right] ,  \label{Hamiltonian}
\end{eqnarray}
where $H_{f,b}^{0}\left( {\bf x}\right) =-\hbar ^{2}\nabla ^{2}/2m_{f,b}+V_{%
\text{trap}}^{f,b}\left( {\bf x}\right) $ and $H_{m}^{0}\left( {\bf x}%
\right) =-\hbar ^{2}\nabla ^{2}/2m_{m}+V_{\text{trap}}^{m}\left( {\bf x}%
\right) +\nu _{\text{bare}}\left( B\right) $ are the Hamiltonians of the
non-interacting gases of fermionic atoms, bosons and molecules,
respectively. The corresponding masses of these atoms and molecules, and the
trapping potentials confining them are indexed by $f,b$\ and $m$\ to
indicate fermionic atoms, bosons and molecules, respectively. In the last
term we stress that the parameter $\nu _{\text{bare}}$ appearing in the
Hamiltonian (\ref{Hamiltonian}) is the bare molecular binding energy \cite
{kokkelmansPRA65}. The true binding energy $\nu \left( B\right) $ given in (%
\ref{nu}) differs from this quantity due to virtual dissociations of the
molecules. Background scattering is omitted, because we assume that the
system is brought close enough to resonance so that the resonant
contribution is much more important.

Starting from Hamiltonian (\ref{Hamiltonian}), we wish to study the dynamics
of the mixture when the scattering length is suddenly brought close to a
Feshbach resonance. In order to grasp the basic physics, some approximations
are made. First of all we assume that there are much more condensate atoms
than fermions, so that the effect on the condensate can be neglected. Also
the depletion of the condensate is presumed to be small, which allows for
the Bogoliubov approximation. Finally we assume that the distance that a
particle travels during a molecule formation process is small compared to
the harmonic oscillator length. The latter permits us to restrict ourselves
to homogeneous systems at first and use a local density approximation to
take the inhomogeneity due to the confinement potential into account at a
later stage. All these assumptions will be justified in Sec. III.

We thus consider a homogeneous boson-fermion mixture without interactions
before $t=0$, the moment when the magnetic field is suddenly brought close
to resonance and the interactions between the atoms start taking place. At $%
t\leqslant 0$, the initial state consists of condensate bosons, `slow'
fermionic atoms (filling the Fermi sphere), and no molecules. From $t=0$
onwards, the evolution of this mixture is described by (\ref{Hamiltonian})$.$
The evolution will consist in the formation of molecules and their
subsequent decay. Pauli blocking will limit the decay channels. However, the
molecular binding energy $\nu $ is, even for magnetic fields that are only $%
0.1%
\mathop{\rm G}%
$ away from resonance, much larger than the Fermi energy (since $\nu =0.1%
\mathop{\rm G}%
\times \mu _{B}/k_{B}\approx 6.7\mu 
\mathop{\rm K}%
$, with $\mu _{B}$ and $k_{B}$ respectively the Bohr magneton and the
Boltzmann constant, whereas $E_{F}$\ is of the order of a few hundred nK)
and blocking effects will be small. As a consequence, a single particle
description of the fermions should be a good first approximation. The
simplified Hamiltonian for this fermion-boson mixture reads 
\begin{eqnarray}
H &=&\int \frac{d^{3}{\bf k}}{\left( 2\pi \right) ^{3}}\left[ \frac{\hbar
^{2}k^{2}}{2m_{f}}\psi _{f}^{\dag }\left( {\bf k}\right) \psi _{f}\left( 
{\bf k}\right) +\frac{\hbar ^{2}k^{2}}{2m_{b}}\psi _{b}^{\dag }\left( {\bf k}%
\right) \psi _{b}\left( {\bf k}\right) +\left( \frac{\hbar ^{2}k^{2}}{2m_{m}}%
+\nu _{\text{bare}}\right) \psi _{m}^{\dag }\left( {\bf k}\right) \psi
_{m}\left( {\bf k}\right) \right]  \nonumber \\
&&+g_{m}\sqrt{n_{b}}\int \frac{d^{3}{\bf k}}{\left( 2\pi \right) ^{3}}\psi
_{m}^{\dag }\left( {\bf k}\right) \psi _{f}\left( {\bf k}\right) +g_{m}\int 
\frac{d^{3}{\bf k}_{f}}{\left( 2\pi \right) ^{3}}\frac{d^{3}{\bf k}_{b}}{%
\left( 2\pi \right) ^{3}}\psi _{m}^{\dag }\left( {\bf k}_{f}+{\bf k}%
_{b}\right) \psi _{f}\left( {\bf k}_{f}\right) \psi _{b}\left( {\bf k}%
_{b}\right) +h.c.  \label{Hsimp}
\end{eqnarray}
Here, $g_{m}\sqrt{n_{b}}$ has the dimension of energy and we will use 
\begin{equation}
\tau =t\times g_{m}\sqrt{n_{b}}/\hbar ,\quad \text{and\quad }\tilde{m}%
=g_{m}m_{r}/(\hbar ^{2}n_{b}^{1/6})  \label{dimless}
\end{equation}
as our dimensionless quantities,\ with $m_{r}=\left(
m_{f}^{-1}+m_{b}^{-1}\right) ^{-1}$\ the reduced mass of a fermion-boson
pair. The first line in (\ref{Hsimp}) represents the free gases of fermionic
atoms, bosons and molecules, respectively. The second line in (\ref{Hsimp})
describes the interaction converting a fermionic atom and a boson into a
molecule (`slow' or `fast') and vice versa. In the single particle picture,
we should study the time evolution of each fermion in the Fermi sea
individually. However, because the Fermi level is much smaller than the
molecular binding energy, the initial kinetic energy of the `slow' fermion
does not influence the final result for the dynamics and can be neglected.
Consequently all the `slow' fermions will have a similar time-evolution due
to the interaction with the condensate, irrespective of the specific value
of their momentum. Thus, it will suffice to analyze the properties of a
fermion with initial momentum $k=0$ to know the evolution of all of them.

An appropriate wave function to describe one fermion that is initially at
rest is 
\begin{eqnarray}
\left| \psi \left( t\right) \right\rangle  &=&[f\left( t\right) \psi
_{f}^{\dag }\left( k_{f}=0\right) \left| 0\right\rangle +m\left( t\right)
\psi _{m}^{\dag }\left( k_{m}=0\right) \left| 0\right\rangle   \nonumber \\
&&+\int d^{3}{\bf k}_{f}\,w\left( k_{f},t\right) \psi _{f}^{\dag }\left( 
{\bf k}_{f}\right) \psi _{b}^{\dag }\left( -{\bf k}_{f}\right) \left|
0\right\rangle ]\times \left| N_{b}\right\rangle ,  \label{psi}
\end{eqnarray}
where $f\left( t\right) $ and $m\left( t\right) $ represent the amplitude of
the bare fermionic atom and the bare molecule. The third term, containing
the amplitude $w\left( k_{f},t\right) $ arises from molecules that
dissociate in a fast fermionic atom and a thermal boson, which is an
important process. Higher order processes where the fast fermionic atom
forms again a molecule that dissociates are neglected in this approach.
Diagrammatically speaking, this comes down to retaining only the diagrams of
Fig. 1\ (a) and neglecting the contributions like those in Fig. 1\ (b).

The Schr\"{o}dinger equation $i\hbar \partial _{t}\left| \psi \right\rangle
=H\left| \psi \right\rangle $\ allows to find the time evolution of the
amplitudes in wave function (\ref{psi}): 
\begin{eqnarray}
i\hbar \frac{d}{dt}f\left( t\right) &=&g_{m}\sqrt{n_{b}}m\left( t\right) ,
\label{dvgl1} \\
i\hbar \frac{d}{dt}m\left( t\right) &=&\nu _{\text{bare}}+g_{m}\sqrt{n_{b}}%
f\left( t\right) +g_{m}\int \frac{d^{3}{\bf k}_{f}}{\left( 2\pi \right) ^{3}}%
w\left( k_{f},t\right) ,  \label{dvgl2} \\
i\hbar \frac{d}{dt}w\left( k_{f},t\right) &=&\frac{\hbar ^{2}k_{f}^{2}}{m_{r}%
}w\left( k_{f},t\right) +g_{m}m\left( t\right) .  \label{dvgl3}
\end{eqnarray}
We choose as initial conditions $f\left( t=0\right) =1$ and $m\left(
t=0\right) =w\left( k_{f},t=0\right) =0$. This corresponds to the initial
state wave function describing one `slow' fermionic atom $\left| \psi
_{init}\right\rangle =\psi _{f}^{\dag }\left( k_{f}=0\right) \left|
0\right\rangle .$ Note that $\sqrt{n_{b}}$\ is assumed time-independent. The
linear differential equations (\ref{dvgl1})-(\ref{dvgl3}) can be solved with
a Laplace transform, yielding 
\begin{eqnarray}
f\left( \tau \right) &=&\frac{1}{\omega _{0}\left( 2\omega _{0}+\tilde{\nu}+%
\frac{3}{8\pi }\left( 2\tilde{m}\right) ^{3/2}\sqrt{\omega _{0}}\right) }%
e^{i\omega _{0}\tau }  \nonumber \\
&&+\int_{0}^{\infty }e^{-i\omega \tau }\frac{1}{4\pi ^{2}}\frac{\left( 2%
\tilde{m}\right) ^{3/2}\sqrt{\omega }}{\left( \omega \left( \omega -\tilde{%
\nu}\right) -1\right) ^{2}+\frac{\left( 2\tilde{m}\omega \right) ^{3}}{16\pi
^{2}}}d\omega ,  \label{ft}
\end{eqnarray}
and 
\begin{eqnarray}
m\left( \tau \right) &=&\frac{-1}{2\omega _{0}+\tilde{\nu}+\frac{3}{8\pi }%
\left( 2\tilde{m}\right) ^{3/2}\sqrt{\omega _{0}}}e^{i\omega _{0}\tau } 
\nonumber \\
&&+\int_{0}^{\infty }e^{-i\omega \tau }\frac{1}{4\pi ^{2}}\frac{\left( 2%
\tilde{m}\omega \right) ^{3/2}}{\left( \omega \left( \omega -\tilde{\nu}%
\right) -1\right) ^{2}+\frac{\left( 2\tilde{m}\omega \right) ^{3}}{16\pi ^{2}%
}}d\omega ,  \label{mt}
\end{eqnarray}
where $\omega _{0}$ is a root of $\omega (\omega +\tilde{\nu}+\frac{1}{4\pi }%
\left( 2\tilde{m}\right) ^{3/2}\sqrt{\omega })-1=0$ and $\tilde{\nu}$\ is a
dimensionless parameter related to the real molecular binding energy $\nu $\
by $\tilde{\nu}=\nu /(g_{m}\sqrt{n_{b}})$. In evaluating the integral in (%
\ref{dvgl2}), we have used the relation between the real and bare molecular
binding energy, $\nu $ and $\nu _{\text{bare}}$, for a momentum cutoff $K$ 
\cite{kokkelmansPRA65}: $\nu =\nu _{\text{bare}}-Kg_{m}^{2}m_{r}/(\pi \hbar
)^{2}.$

\section{Results and discussion}

\subsection{Atom - molecule oscillations}

It is instructive for the understanding of the time evolution to investigate
the eigenstates of the Hamiltonian (\ref{Hsimp}) describing the system in
which the time evolution takes place. The ground state for a single
fermionic atom in a system described by (\ref{Hsimp}) can be investigated by
using$\ $(\ref{psi}) as a variational wave function with $f,m,w(k_{f})$\ as
variational parameters.\ The variational expression for the ground state
energy is 
\begin{eqnarray}
\frac{E_{0}(f,m,w(k_{f}))}{g_{m}\sqrt{n_{b}}} &=&\left( f^{\ast }m+m^{\ast
}f\right) +\int \frac{d^{3}{\bf k}_{f}}{\left( 2\pi \right) ^{3}}\frac{%
k_{f}^{2}}{2\tilde{m}}\left| w\left( k_{f}\right) \right| ^{2}  \nonumber \\
&&+\int \frac{d^{3}{\bf k}_{f}}{\left( 2\pi \right) ^{3}}\left( m^{\ast
}w\left( k_{f}\right) +w^{\ast }\left( k_{f}\right) m\right) ,
\end{eqnarray}
where $k_{f}$ is now measured in units $n_{b}^{1/3}.$ Minimizing this
expression, with the constraint $\left| \left\langle \psi |\psi
\right\rangle \right| =1$ we find for the ground state energy the equation 
\begin{equation}
\tilde{E}\left( \tilde{E}-\tilde{\nu}-\int \frac{d^{3}{\bf k}_{f}}{\left(
2\pi \right) ^{3}}\frac{1}{\tilde{E}-\frac{k_{f}^{2}}{2\tilde{m}}}\right)
-1=0,  \label{Evareq}
\end{equation}
where $\tilde{E}=E/(g_{m}\sqrt{n_{b}}),$ which has as lowest solution
(relative to the bottom of the fermion band) $\tilde{E}=-\omega _{0}$. The
ground state is a bound state representing a coherent superposition of the
states corresponding to the fermion as an atom and as a part in a molecule: 
\begin{eqnarray}
\left| \psi _{bound}\right\rangle ={\cal N} &\,&\left[ \frac{-1}{\omega _{0}}%
\psi _{f}^{\dag }\left( k_{f}=0\right) +\psi _{m}^{\dag }\left(
k_{m}=0\right) \right. \\
&&+\left. \int \frac{d^{3}{\bf k}_{f}}{\left( 2\pi \right) ^{3}}\frac{1}{%
-\omega _{0}-\frac{k_{f}^{2}}{2\tilde{m}}}\psi _{f}^{\dag }\left( {\bf k}%
_{f}\right) \psi _{b}^{\dag }\left( -{\bf k}_{f}\right) \right] \left|
0\right\rangle ,  \nonumber
\end{eqnarray}
with ${\cal N}$ a normalization constant. Similar bound states are found in
the photodissociation of a negative ion \cite{RzazewskiJPhysB15}. Equation (%
\ref{Evareq}) has only one solution for $E$ negative. The solutions of (\ref
{Evareq}) with $\tilde{E}=\omega >0$ form a continuum, because we consider
the limit of an infinite system. The eigenstates $\left| \psi _{exc}\left(
\omega \right) \right\rangle $ corresponding to these solutions are the
excited states of the Hamiltonian (\ref{Hsimp}).

The time evolution of the initial state $\left| \psi _{init}\right\rangle $\
can then be written in the basis of the eigenstates of (\ref{Hsimp}) as 
\begin{eqnarray}
\left| \psi (t)\right\rangle  &=&\left| \psi _{bound}\right\rangle
e^{i\omega _{0}t}\left\langle \psi _{bound}|\psi _{init}\right\rangle  
\nonumber \\
&&+\int_{0}^{\infty }\left| \psi _{exc}(\omega )\right\rangle e^{-i\omega
t}\left\langle \psi _{exc}(\omega )|\psi _{init}\right\rangle d\omega .
\label{psivant}
\end{eqnarray}
In Fig. 2, the modulus square of $\left\langle \psi _{bound}|\psi
_{init}\right\rangle $,\ is shown for several values of $\tilde{\nu}$ in the
left panel. In the right panel, the modulus square of the projection of the
bound state on the molecular state, $\left| \left\langle \psi
_{bound}\right| \psi _{m}^{\dag }\left( k_{m}=0\right) \left| 0\right\rangle
\right| ^{2}$,\ is shown. Note that expression (\ref{psivant}) has the same
structure as expressions (\ref{ft}),(\ref{mt}). This is to be expected,
because the amplitudes $f(t)$\ and $m(t)$ represent the projection of $%
\left| \psi (t)\right\rangle $\ on the `slow' fermionic atom component and
the `slow' molecule component: 
\begin{eqnarray*}
f(t) &=&\left\langle 0\right| \psi _{f}\left( k_{f}=0\right) \left| \psi
(t)\right\rangle , \\
m(t) &=&\left\langle 0\right| \psi _{m}\left( k_{f}=0\right) \left| \psi
(t)\right\rangle .
\end{eqnarray*}
The two terms appearing in $\left| \psi (t)\right\rangle ,$ expression (\ref
{psivant})\ give rise to the corresponding two terms in $f(t)$, expression (%
\ref{ft}), and in $m(t)$, expression (\ref{mt}). The first term (the first
line in expressions\ (\ref{ft}),(\ref{mt})) is related to the bound state
and the second term (containing the integration over $\omega $) is related
to the excited states. Because of the continuous character of the excited
states as a function of $\omega $, the terms related to the excited states
vanish for large times.

We find that for large $\left| \tilde{\nu}\right| $, the conversion from
`slow' to `fast' fermions is greatly reduced. This can be understood from
the fact that for large $\left| \tilde{\nu}\right| $, $\left\langle \psi
_{bound}|\psi _{init}\right\rangle $ becomes very small (as can be seen in
Fig. 2) and the initial state coincides nearly with an eigenstate of the
system, $\left| \psi _{init}\right\rangle \approx \left| \psi _{exc}(\omega
=0)\right\rangle $ so that $\left| \psi (t)\right\rangle \approx \left| \psi
_{init}\right\rangle $ and $f(t)$ remains close to 1 (its initial value).

The fraction of `slow' molecules or `slow' fermionic atoms, obtained from (%
\ref{ft}) and (\ref{mt}) respectively, displays damped oscillations
resulting from the interference between the terms related to the bound state
and the terms related to the excited states. Fig. 3 shows the evolution of
the densities of `slow' fermionic atoms and `slow' molecules for different
values of the molecular binding energy $\tilde{\nu}$ and mass $\tilde{m}$.
For the chosen values of the mass, the oscillations are visible for negative
molecular binding energy $\tilde{\nu}<0$, while for $\tilde{\nu}$ positive
they are strongly damped.

We observed that the larger $\tilde{m},$ the fewer molecules are formed.
Indeed, according to Fermi's golden rule, a bigger reduced mass means a
faster dissociation rate ($\sim m_{r}^{3/2}$) of the of molecules.
Remarkably, also for $\tilde{m}\gg 1$, the fraction of `slow' fermions
reduces substantially over time even though at all times few molecules that
can dissociate to `fast' fermionic atoms are present. This effect can be
understood since, even though few molecules are present at any time, they
form and dissociate quickly. This raises the question whether our
approximate wave function (\ref{psi}) is good enough. For low $\tilde{m}$,
it is obvious that the dissociation of molecules will be slow and strong
Rabi oscillations between fermionic atoms and molecules occur. The third
term of (\ref{psi}), containing the factor $w(k_{f},t)${\bf ,} is then
relatively unimportant and corrections on it will not affect the result a
lot. For high $\tilde{m}$, the molecules dissociate rapidly and the
contribution to $\left| \psi \left( t\right) \right\rangle $ coming from the
pairs of `fast' fermionic atoms and bosons (last term from equation (\ref
{psi})) must be big. But in this case the corrections to the contribution of
these pairs --through new formation of fast molecules out of a fast
fermionic atom-- are small, because few of those molecules are formed.
Diagrammatically, this means that diagrams such as those in Fig.\ 1\ (b) are
negligible. Then for intermediate $\tilde{m}$ one can also expect (\ref{psi}%
) to be a good approximation, at least in order to calculate the fraction of
`slow' fermions.

\subsection{Relation to experiments on $^{40}$K-$^{87}$Rb mixtures}

To check whether the picture described here, with fermions that form a bound
state with the condensate, is correct, one has to be able to measure
preferably the fraction of `slow' fermionic atoms. This should, after some
transient behavior, go to a constant value and not to zero. They can be
detected by a time of flight expansion after a magnetic pulse has been
applied.

In $^{40}$K-$^{87}$Rb plenty of Feshbach resonances have been predicted \cite
{Simoni,SimoniPrivate} allowing us to estimate quantitatively the magnitude
of the model parameters. Therefore, we make use of $g_{m}=\sqrt{2\pi \hbar
^{2}a_{bg}\Delta B\,\Delta \mu /m_{r}}$ , with $a_{bg}$ the background
scattering length. As an example we take the resonance between the $\left(
9/2,-9/2\right) +\left( 1,1\right) $ hyperfine states around $741.2%
\mathop{\rm G}%
$. For $n_{b}=10^{14}%
\mathop{\rm cm}%
^{-3},$ the frequency $g_{m}\sqrt{n_{b}}/\hbar $ is\ equal to $7.72\times
10^{5}%
\mathop{\rm Hz}%
$, which is much larger than the harmonic oscillator frequency. Thus we can
expect that the confinement is relatively unimportant for the atom-molecule
oscillations. We find in this case $\tilde{\nu}=33.13\times \left(
B-B_{0}\right) \left( 
\mathop{\rm G}%
\right) /\sqrt{n_{b}\left( 10^{14}%
\mathop{\rm cm}%
^{-3}\right) }$ and further $\tilde{m}=15.47.$ The values used for the
parameters of Fig. 3 are for this resonance with $n_{b}=10^{14}%
\mathop{\rm cm}%
^{-3}$ and $B-B_{0}=\pm 0.1%
\mathop{\rm G}%
$ on the left, and on the right for the $\left( 9/2,-7/2\right) +\left(
1,1\right) $ hyperfine states with a Feshbach resonance around $522.9%
\mathop{\rm G}%
$.

The current results have been derived in the framework of a translation
invariant system whereas in the experimental setup the inhomogeneity of the
trapped gasses is apparent. The time evolution has been calculated for
homogeneous mixtures, and it is necessary to estimate what the effects due
to the confinement are. Considering the case with much more bosons than
fermions, we can neglect in a first approximation the change of the boson
density. Then one can find in a mean-field approximation the redistribution
speed of the fermions in the trap. The local density approximation is
applicable only if during the time scale of a Rabi oscillation between the
fermionic atom state and the molecular state, the fermions have not moved
over a substantial fraction of the trap length. In the Thomas-Fermi
approximation for the bosons, the mean field potential felt by a fermion is $%
2\pi \hbar ^{2}a_{\text{scatt}}n_{b}^{0}/m_{r}\times \left(
1-r^{2}/R_{C}^{2}\right) ,$ with $n_{b}^{0}$ the condensate density in the
origin and $R_{C}$ the radius of the condensate. This results in an
additional confinement potential (inverted for $a_{\text{scatt}}>0$) acting
on the fermions. For the resonance used above, for $n_{b}^{0}=10^{14}%
\mathop{\rm cm}%
^{-3}$ and $R_{C}=10$ $\mu $m and making use of (\ref{ascatt}), its
frequency equals $\omega _{bf}=4.95\times 10^{3}%
\mathop{\rm s}%
^{-1}/\sqrt{\tilde{\nu}}.$\ For $\tilde{\nu}$ not too small, this is much
lower than the frequency scale of the molecule formation $g_{m}\sqrt{%
n_{b}^{0}}/\hbar =7.72\times 10^{5}%
\mathop{\rm Hz}%
,$ which means that the dynamics due to molecule formation is fast compared
to the mean-field collapse, and the local density approximation can be used.
For $\tilde{\nu}=1$, an atom migrates 10\% of the trap length in 0.75 ms
whereas the period of a Rabi oscillation is around 1.3 $\mu $s$.$

Within the approximations described above, we plot in Fig. 4 the density of
slow fermions as a function of time and distance from the center of the
trapping potential in the case that the fermion cloud has a radius 1.5 times
bigger than the condensate. To use the formalism presented above, space was
subdivided in discrete boxes in which the gases were presumed homogeneous,
as in the local density approximation. As discussed above, no redistribution
of fermions across volume elements is considered so that each volume element
can be treated independently of the others. In the lower panels, the total
fraction, averaged over space, of `slow' fermionic atoms is compared to the
local fraction of `slow' fermionic atoms in the center of the trap. One can
see that the inhomogeneity decreases the amplitude of the oscillations in
the total fraction of `slow' fermionic atoms. The reason is that the
condensate is the most dense in the middle, which increases the efficiency
of the atom-molecule conversion, and also because $\tau ,\tilde{m}$ and $%
\tilde{\nu}$ depend on $n_{b}\left( r\right) ,$ so that on different places
in the condensate oscillations with different frequencies occur.

\section{Conclusions}

We have made a study of the dynamics of a fermion-boson mixture resulting
from a fast switch in the magnetic field, bringing it close to an
interspecies Feshbach resonance. Our analysis shows that damped oscillations
between fermionic atoms and molecules will occur on the typical time scale
of $\hbar /(g_{m}\sqrt{n_{b}})$ ($\approx 1\mu $s for a $^{40}$K-$^{87}$Rb
mixture)$.$ Such a type of oscillations have recently been observed in
single component BEC's near a Feshbach resonance \cite{DonleyNature417}. The
decay of the molecules in pairs of `fast' fermionic atoms and bosons leads
to heating of the system, but a core of `slow' fermions will remain, due to
the existence of a bound state of the fermions coupled to the condensate and
the continuum of thermal bosons. Using the latest data on resonances in $%
^{40}$K-$^{87}$Rb scattering \cite{Simoni,SimoniPrivate}, we find that the
predicted oscillations between slow fermions and slow molecules should be
observable both in a homogeneous and inhomogeneous system. The
redistribution of the densities are expected to occur much slower than these
oscillations.

\section{Acknowledgments}

The authors like to thank G. Modugno for the useful and stimulating
discussions and A. Simoni for providing us the latest theoretical
predictions on Feshbach resonances in $^{40}$K-$^{87}$Rb. Discussions with
M. Inguscio are also gratefully acknowledged. Two of the authors (M. W. and
J. T.) are supported financially by the Fund for Scientific Research -
Flanders (Fonds voor Wetenschappelijk Onderzoek -- Vlaanderen). This
research has been supported financially by the FWO-V projects Nos.
G.0435.03, G.0306.00, the W.O.G. project WO.025.99N.and the GOA BOF UA 2000,
IUAP.

\section*{Figure captions}

FIG. 1: Feynman diagrams that describe the correction to the Green function
of the heteronuclear molecules in the mixture. The terms in the left part
(a) are taken into account in the theory of this paper. Multiple
dissociations as depicted in the right part (b) are neglected.

FIG. 2: Number of molecules (right) and fermionic atoms (left) for $%
t\rightarrow \infty $ as a function of the molecular binding energy $\tilde{%
\nu}$ for different values of the mass $\tilde{m}$.

FIG. 3: Time evolution of the densities of `slow' fermionic atoms and
molecules in a homogeneous system for two different values of the molecular
binding energy and the mass. The molecular binding energies correspond to
detunings of the magnetic field of $B-B_{0}=\pm 0.1\mathop{\rm G}$. For $%
n_{b}=10^{14}\mathop{\rm cm}^{-3}$ the dimensionless detunings and the
masses correspond then to the predicted $^{40}$K-$^{87}$Rb resonances in the 
$\left( 9/2,-9/2\right) +\left( 1,1\right) $ hyperfinestates at $741.2 %
\mathop{\rm G}$ (left) and to the $\left( 9/2,-7/2\right) +\left( 1,1\right) 
$ at $522.9\mathop{\rm G}$ (right).

FIG. 4: Time evolution of the density of `slow' fermionic atoms when a
noninteracting boson-fermion mixture in a harmonic trap is suddenly brought
close to a Feshbach resonance. In the upper panels, the density of `slow'
fermionic atoms is shown as a function of the distance to the center of the
trap and in the lower panels, the spatially averaged fraction of `slow'
fermionic atoms is compared with this fraction in the origin. The
inhomogeneity decreases the spatially averaged oscillations because the
amplitude and frequency depend on the condensate density and consequently
vary in space. The parameters used are the same as for the right panels in
Fig. 3$,$ $\tilde{m}=3.8$ and $\tilde{\nu}=\pm 2.5,$ and correspond to a
realistic situation in $^{40}$K-$^{87}$Rb mixtures.

\end{document}